# Spin light-emitting devices in a 2D magnet

Fanglu Qin[§], Haiyang Liu[§], Aosai Yang, Yilin Liu, Xuanji Wang, Yue Sun, Xinyi Zhou, Zdenek Sofer, Jiayuan Zhou, Xue Liu, Sheng Liu[*], Vanessa Li Zhang, Xiaoze Liu[*], Weibo Gao, Ting Yu[*]

AUTHOR INFORMATION

**Corresponding Author**

Ting Yu − School of Physics and Technology, Wuhan University, Wuhan 430072, China. Wuhan Institute of Quantum Technology, Wuhan 430206, China; orcid.org/0000-0002-0113-2895. Email: yu.ting@whu.edu.cn

Sheng Liu, Xiaoze Liu − School of Physics and Technology, Wuhan University, Wuhan, 430072, P. R. China; Email: liu.sheng@whu.edu.cn, xiaozeliu@whu.edu.cn

**Authors**

Fanglu Qin, Haiyang Liu, Aosai Yang, Yilin Liu, Xuanji Wang, Yue Sun, Xinyi Zhou, Vanessa Li Zhang − School of Physics and Technology, Wuhan University, Wuhan 430072, China.

Zdenek Sofer − Department of Inorganic Chemistry, Faculty of Chemical Technology, University of Chemistry and Technology Prague, Technicka 5, Prague 6 166 28, Czech Republic

Jiayuan Zhou, Xue Liu − Center of Free Electron Laser & High Magnetic Field, Anhui University, Hefei 230601, China

Weibo Gao − Division of Physics and Applied Physics, School of Physical and Mathematical Sciences, Nanyang Technological University, Singapore, 637371 Singapore. School of Electrical and Electronic Engineering, Nanyang Technological University, Singapore, 639798 Singapore. Centre for Quantum Technologies, Nanyang Technological University, Singapore, 117543 Singapore

**Abstract**

Emerging two-dimensional (2D) magnetic semiconductors represent transformative platforms to explore magneto-optics and opto-spintronic applications. Though 2D opto-spintronics has attracted tremendous research efforts in spin-dependent photodetectors and non-volatile memory components, the realization of one core application - spin-modulated light-emitting device (spin-LED) - remains elusive so far. Here we successfully realize prototype spin-LED integrated with a 2D semiconducting magnet CrSBr, demonstrating considerable electroluminescence (EL) down to bilayers. Intriguingly, the EL of the spin-LED is discovered to be directly manipulated by spin-flip and spin-canting transitions. Notably, spin-flip transitions enable unprecedented hysteretic behaviors of EL characteristics, while spin-canting transitions induce EL continuous modulation with robust anisotropy. This versatile manipulation is originated from the synergy of magnetic-order mediated excitonic transitions and spintronic transport. The prototype demonstration of spin-LED establishes an indispensable scheme of opto-spintronic devices leveraging 2D spin transitions and strong excitonic effects, presenting a critical step towards integrated 2D opto-spintronics.

# Introduction

Opto-spintronics, merging photonic, electronic, and magnetic functionalities, becomes a novel but desirable paradigm for future information technologies[1-4]. Compared with conventional charge-based electronics, this paradigm exploits electron spins as the fundamental variable for information processing and storage, promising new functionalities with significantly reduced energy consumption. For instance, non-volatile memory devices feature optical writing and electrical readout of the spin orders[5,6], and neuromorphic computing hardware employ spin-based artificial synapses[7]. By leveraging spin-wave excitations, opto-magnonic circuits and quantum transduction platforms have also been proposed and developed[8]. As one of the fundamental optoelectronic components, light-emitting devices (LEDs) advance to unexplored versatility when their electrical injections involve profound spin manipulations[9-12].

The emerging vdW magnets have been recently extensively studied for striking fundamental physical breakthroughs with unconventional two-dimensional (2D) magnetism[13,14]. The confluence of inherent 2D magnetism, electronic structures, and weak dielectric screening in the layered structures suggests unprecedented functionalities to opto-spintronic components[9]. For example, spin-dependent photodetectors[15-17], spin-photovoltaics[18] with magnetic $CrI_3$ and MnSe have been proposed; Moreover, non-volatile memory component with magnetic four-layer $CrI_3$ and $VSe_2$ has also been designed[19]. However, as one core application of 2D opto-spintronics, no LED with spin-based functionality has been realized so far[20-23]. Here, we report the realization of a prototype spin-modulated LED (spin-LED) in CrSBr-based quantum well (QW) structures, achieving considerable electroluminescence (EL) down to bilayer CrSBr. Unlike EL in conventional semiconductors dictated by band structure and carrier recombinations[24-27], the spin-exciton coupling in CrSBr introduces versatile manipulations of its EL. Remarkably, the EL can be fully manipulated with pronounced magnetic hysteresis by spin-flip transitions; Meanwhile, the EL can be modulated with spin-canting features by perpendicular fields with respect to the easy axis. Moreover, the modulation of EL efficiency during these two spin transitions shows the opposite trend, due to the distinct magneto-resistance responses in these two cases.

In this prototype spin-LED, the air-stable vdW CrSBr is selected as the magnetic medium with layer-dependent magnetism including ferromagnetic (FM), Ferrimagnetic (FiM) and antiferromagnetic (AFM) phases[28-32]. This selection is deliberated based on the unique exciton/electron-spin interactions in CrSBr as elucidated in recent reports[28]. Due to the 2D quantum confinements, strong anisotropy and 2D magnetism, CrSBr exhibits unconventional light-matter interactions for photoluminescence (PL)[33], surface and bulk excitons[34,35], exciton polaritons[36-38], exciton-phonon/exciton-magnon coupling[39,40] and unique exciton dynamics[41]. Moreover, CrSBr also supports distinct spin-charge coupling for spin-dependent charge transfer[42] and electrostatic control of spin polarization in heterostructures[31]. This deliberate selection of CrSBr gives rise to the first prototype opto-spintronic devices of LED modulated by spin transitions, and promises the feasibility of all-2D opto-spintronics.

## EL characterization in CrSBr

**Sample configuration and characterizations.** To probe the intrinsic EL properties of CrSBr, a tunneling LED integrated with a QW structure is designed and fabricated (see Methods and Supplementary Fig. S1, S2 for sample preparations). The QW structure consists of a few-layer CrSBr flake sandwiched between top and bottom graphite (G) electrodes, with upper and lower barrier layers of hexagonal boron nitride (h-BN) (Fig. 1a). In this device design, electrons and holes are injected into the CrSBr flake by tunneling through hBN from top and bottom G, respectively (Fig. 1b). This design is crucial as it provides clean interfaces and minimizes defect-induced scattering, both essential for achieving high-quality optical and electrical performance[43-45]. As a result, few-layer CrSBr samples not only show clear PL spectra of bulk ($X_b$) and surface ($X_s$) excitons, but also more importantly exhibit unambiguous EL spectra as well. Fig. 1c shows the CrSBr thickness-dependent spectral features of $X_b$ and $X_s$ in the device structure, which are consistent with the previously reported excitonic emissions in CrSBr[34]. Specifically, the PL features only one peak of $X_s \sim 1.34$ eV (the blue segment) in the bilayer (2L) CrSBr, and starts to embrace another peak of $X_b \sim 1.36$ eV (pink segment) in samples with 3-layer (3L) and more layers of CrSBr (Supplementary Fig. S3). As the CrSBr thickness increases to 15 layers (15L), the PL spectrum is dominated by the peak of $X_b$. Note here two split peaks of $X_s$ in the 15L sample may be due to two $X_s$ experiencing slightly different dielectric environments[34,35]. Expectedly, this thickness dependence originates from the layered structure and the AFM orders of CrSBr, making $X_b$ highly confined to each individual layer and $X_s$ extremely sensitive to the thickness and dielectric environments[34].

**Observation of EL.** With the PL characterizations in these CrSBr LED samples, the EL has been successfully observed with prominent thickness dependence (Fig. 1d). Apparently, all the EL spectra are dominated by $X_s$, where its peaks are consistent with the PL spectra but their full-width-at-half-maximum (FWHM) becomes much narrower probably due to the reduced screening by electric injection and suppressed non-radiative decay channel. As the thickness of CrSBr increases, the peaks of $X_s$ also split into two being consistent with the PL. Moreover, statistical analysis of thickness-dependent spectra reveals that the $X_s$ emission intensity scales linearly with layer number (2–15 layers) in both EL and PL measurements (Supplementary Fig. S4) This linear dependence suggests the excitons at each layer of CrSBr are more likely to recombine independently with quantum confinements induced by the AFM mediated interlayer interactions[34]. More importantly, the consistency of EL infers that LED device structure does not compromise the AFM mediated interlayer interactions. However, the EL peak of $X_b$ is much weaker than the PL at thicker samples even though the peak positions of $X_b$ are consistent (Fig. 1d insets). This is most likely due to efficient recombination of the injected carriers at the surface, but reduced recombination of much less injected carriers in the bulk region.

To further probe the EL characteristics, voltage dependence and polarization dependence are both investigated. The voltage dependence unambiguously demonstrates the tunneling features of EL, verifying the high quality of the devices and its critical role in the realization of EL. Fig. 1e demonstrates a sharp increase in EL intensity of 15-layer CrSBr as the applied bias rises from 2 V to 4.5 V at 2 K, indicating clear tunneling behavior. Fitting analysis of this voltage-dependent EL reveals the specific trend of nonlinear intensity enhancement (Supplementary Information Section II, Fig. S5). Temperature-dependent EL measurements further highlight the excellent contact of the QW device, which exhibits near-ohmic behavior at 2 K, with the EL intensity increasing linearly as the temperature decreases from 70 K to 2 K (see Supplementary Information Section II, Fig. S6).

The polarization dependence of EL indicate robust anisotropy, which is resilient to external magnetic field ($B$) and has little dependence on CrSBr thicknesses. As shown in Fig. 1f, the EL of the 15-layer CrSBr device maintains a strong linear polarization that remains invariant under both in-plane magnetic field $B_{in}$ of 0.5 T along the easy axis of $b$ and out-of-plane field $B_{out}$ (3 T) along the hard axis of $c$. This is consistent with the near-unity linear polarization of PL. Further calculations of the EL linear polarization ratio $\rho=\frac{I_0-I_{90}}{I_0+I_{90}}$ reveal that ρ approaches 100% under 0 T, $B_{in}$ (0.5 T), and $B_{out}$ (3 T) (Supplementary Information Section II, Fig. S7). Though the EL linear polarization might be correlated with the in-plane spin orders[26], the in-plane anisotropy is not broken down by the applied magnetic fields. Moreover, similar EL linear polarization is also observed in 9-layer CrSBr device (see Supplementary Information Section II, Fig. S8), and thus the anisotropy is intrinsic and has little dependence on CrSBr thicknesses. This property unlocks applications in polarized optoelectronics, quantum photonics, and magneto-optic sensing—all benefiting from the decoupling of spin and optical anisotropy[46,47].

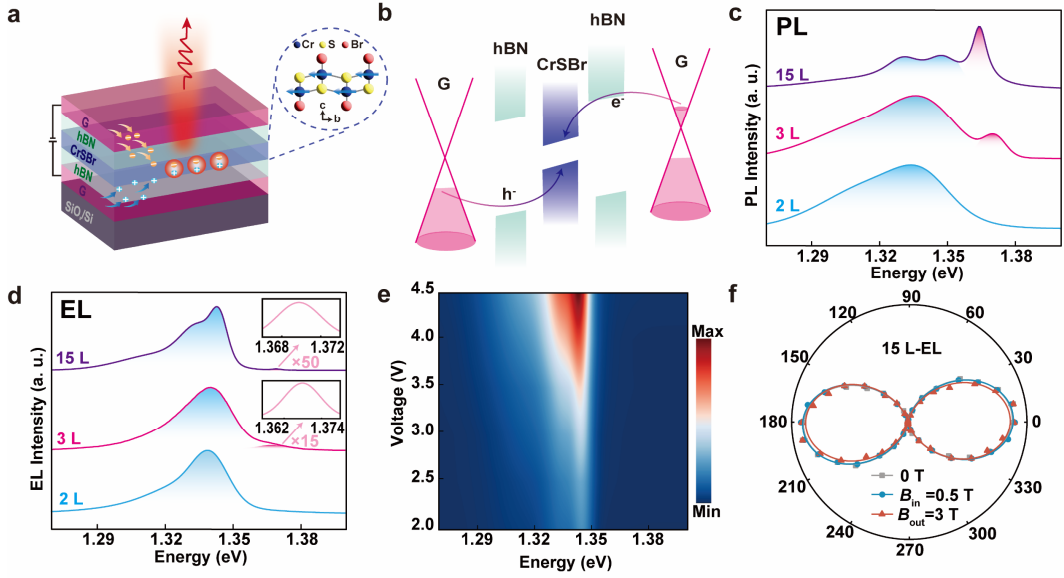

**Fig. 1 EL characterization of CrSBr**. (**a**) Cross-sectional illustration of the engineered van der Waals heterostructure with vertically assembled components: G/h-BN/CrSBr/h-BN/G. Magnified view showing the crystal structure and magnetic configuration of CrSBr. Blue, yellow and red balls represent Cr, S and Br atoms, respectively. (**b**) Schematic of tunneling mechanism in a CrSBr-based QW. PL (**c**) and EL (**d**) spectra of 2-, 3- and 15-layer CrSBr. Insets in (d) display an enlarged view of the $X_b$ constituent. (**e**) EL spectra of 15-layer CrSBr acquired under applied bias voltages between 2 V and 4.5 V at 2 K. (**f**) Linear polarization of EL as polar plot for 15-layer CrSBr under 0 T, 0.5 T (in-plane), 3 T (out-of-plane) $B$. Experimental data are represented by dots; solid curves correspond to fitting results.

## EL manipulation by spin-flip transitions

**EL hysteresis of peak energies**. To monitor the EL by spin-flip transitions, the EL dependence on the $B$ along the magnetic easy axis of CrSBr ($B_{in} \parallel b$) is firstly investigated in an LED device of 15L CrSBr at 2 K. The evolution of EL spectra is summarized as the field $B_{in} \parallel b$ is swept backward from 0.6 T to -0.6 T (Fig. 2a) and forward from -0.6 T to 0.6 T (Fig. 2b). When $B_{in} \parallel b$ is swept to stronger fields around -0.3T and 0.3T, the EL peaks of both $X_s$ and $X_b$ undergo abrupt redshifts by ~ 20 meV. The redshifts of CrSBr exciton energies are recognized as an effective indicator for the spin transitions from AFM to FM, where the excitons of FM phase are smaller than those of AFM phase by ~ 20 meV[28,34]. Therefore, the abrupt redshifts of ~ 20 meV are most likely to represent the spin-flip transitions between AFM and FM phases along the easy *b* axis. More strikingly, by extracting the *B*-dependent peak energies of $X_s$ for the sweeping loop, the EL peaks demonstrate a pronounced hysteresis of the abrupt redshifts (Fig. 2c, see a similar hysteresis for the EL peak of $X_b$ in Supplementary Fig. S9). The EL hysteresis of abrupt redshifts is thus smoking-gun evidence to indicate the underlying spin-flip transitions in the operating LED devices.

To further confirm the spin-flip transitions, the PL spectra of the same device were also measured under sweeping $B_{in} \parallel b$. The PL spectra show consistent abrupt excitonic redshifts around -0.3 T and 0.3 T (Figs. 2d and 2e), and the *B*-dependent PL peaks also demonstrate a similar hysteresis (Fig. 2f). Furthermore, both the EL and PL spectral changes via spin-flip transitions persist throughout various CrSBr thicknesses, though the magnitude of the abrupt redshifts diminish progressively with decreasing layers (see Supplementary Information Section III, Fig. S10-S13 for 9 L and 3 L CrSBr devices). The PL hysteresis thus corroborates the spin-transition mediated exciton energies. It is also noted that the EL hysteresis is more prominent than that of PL. This might be due to the enhancement of the itinerant ferromagnetic component from the carrier injection[48], while the PL may involve more extrinsic effects such as stronger scattering under laser illumination and partially masking the intrinsic magnetic response for hyesteresis[49].

**EL hysteresis of intensities.** Other than the EL peak energies, the EL intensities also show intriguing hysteresis under sweeping $B_{in} \parallel b$, which is most probably due to the hysteresis magnetoresistance (MR) in the devices. Fig. 2g summarizes the MR measurements of few-layer CrSBr under sweeping $B_{in} \parallel b$, where the MR decreases with increasing *B* magnitude and reaches the minimum around -0.3 T and 0.3 T without further changes. By tracking the full sweeping loop of $B_{in} \parallel b$, the MR shows a hysteresis feature which is inherited from the hysteresis of the magnetic phases. In the magnetic hysteresis, MR is lower in parallel alignment of magnetic moments in the FM phase, but becomes larger in antiparallel alignment in the AFM phase[48,49]. Therefore, this MR hysteresis could directly manipulate the EL intensities in the studied LED devices.

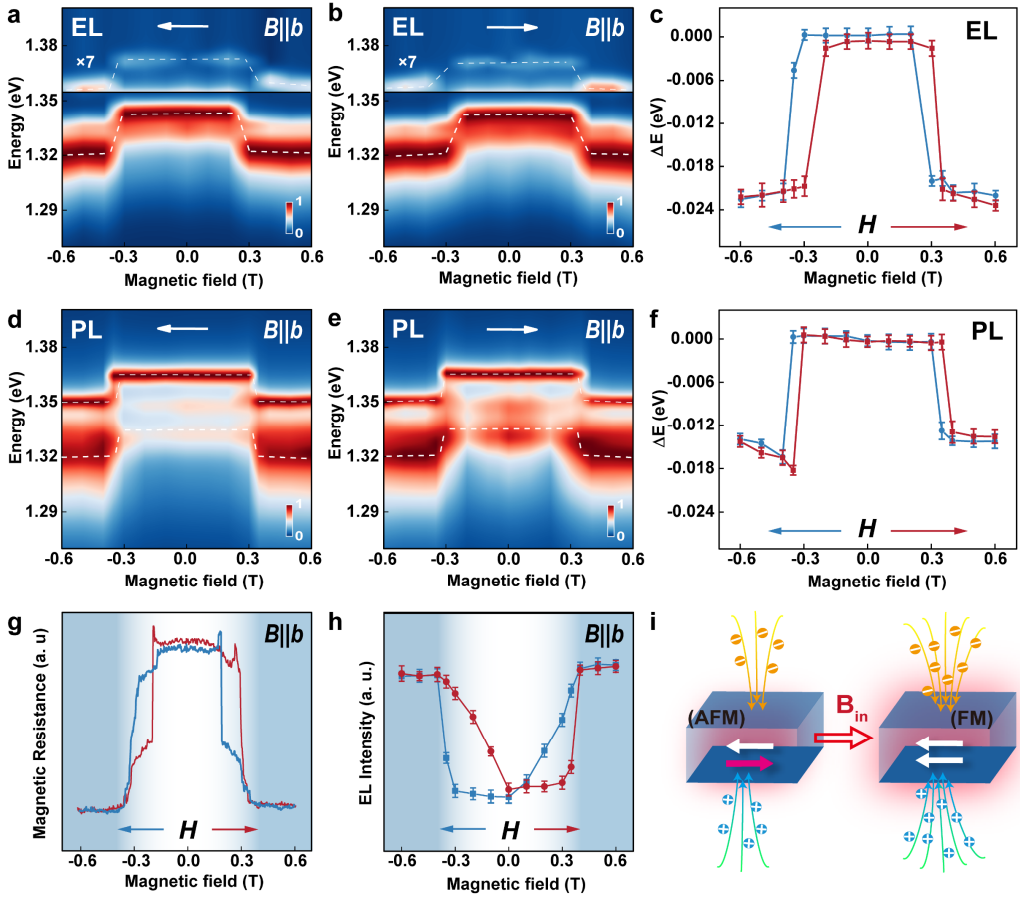

**Fig. 2 EL manipulation by spin-flip transitions.** Normalized EL spectra of CrSBr as $B$ is swept along the easy axis from 0.6 T to -0.6 T (**a**) and from -0.6 T to 0.6 T (**b**). White dashed lines guide the eye to the evolution of $X_s$ and $X_b$ with $B$. (**c**) Difference in EL energy shift of $X_s$ during a back-and-forth sweep of $B$, where 0 T is referenced to 0 eV. PL spectra of CrSBr as $B$ is swept along the easy axis from 0.6 T to -0.6 T (**d**) and from -0.6 T to 0.6 T (**e**). White dashed lines guide the eye to the evolution of $X_s$ and $X_b$ with $B$. (**f**) Calculation of the energy shift difference for the PL spectra during a back-and-forth sweep of $B$, with 0 T referenced to 0 eV. (**g**) MR of CrSBr as $B$ is swept along easy axis. (**h**) EL intensity of CrSBr as $B$ is swept along easy axis. (**i**) The mechanism of in-plane-$B$ modulated EL emission.

More surprisingly, the EL intensities demonstrate a similar hysteresis of the $X_s$ intensities with the sweeping loop of $B_{in}$ ∥ $b$ in the 15 L device, where the intensities are inverse proportional to the MR (Fig. 2h). The AFM phase, which corresponds to higher MR, results in weak EL intensity. After the spin-flip transition to the FM phase, the lower MR facilitates more efficient carrier injection, yielding stronger EL intensity. Further analysis shows that the modulation efficiency of EL by $B_{in}$ reaches 48.9%, as discussed in detail in Supplementary Information Section III. To illustrate the underlying mechanism of the EL hysteresis, the mechanism schematics is also presented in Fig. 2i to provide a physical intuition of the spin-order correlated EL intensities.

## EL manipulation by spin-canting transitions

**Continuous modulation of EL peak energies.** To monitor the EL by spin-canting transitions, the EL dependence on the out-of-plane *B* along the magnetic hard axis of CrSBr ($B_{out}$ ∥ $c$) is investigated in the device of 15L CrSBr at 2 K. The evolution of EL spectra is summarized as the field $B_{out}$ ∥ $c$ is swept backward from 3.0 T to -3.0 T (Fig. 3a) and forward from -3.0 T to 3.0 T (Fig. 3b). When $B_{out}$ ∥ $c$ is swept to stronger fields around -2.0 T and 2.0 T, the EL peaks of $X_s$ and $X_b$ undergo continuous redshifts, which are saturated beyond higher magnetic fields of ~2.0 T or ~ -2.0 T. By extracting the *B*-dependent peak energies of $X_s$ and $X_b$ for the sweeping loop, the redshifts show a general spin-canting feature following ($\Delta E \propto B_{out}^2$) (Fig. 3c). For comparisons, the PL spectra under the sweep loop of $B_{out}$ ∥ $c$ were also measured, showing consistent continuous redshifts between -2.0 T and 2.0 T (Figs. 3d and 3e). The $B_{out}$-dependent PL peaks also demonstrate a similar spin-canting feature (Fig. 3f), which is consistent with previous spin-canting observation[28] and corroborates the general feature of spin-canting modulation for both $X_s$ and $X_b$. Moreover, this spin-canting response of EL is consistently observed across 9-layer and 3-layer CrSBr (Supplementary Information Section IV, Fig. S14 - S17), confirming it as an intrinsic characteristic of the EL manipulation. It is noted that there is a small plateau of the $B_{out}$-dependent EL peak ($X_s$) energies at small magnetic fields (Fig. 3c bottom), which may be due to spin canting barriers at the interface during the electrical injection.

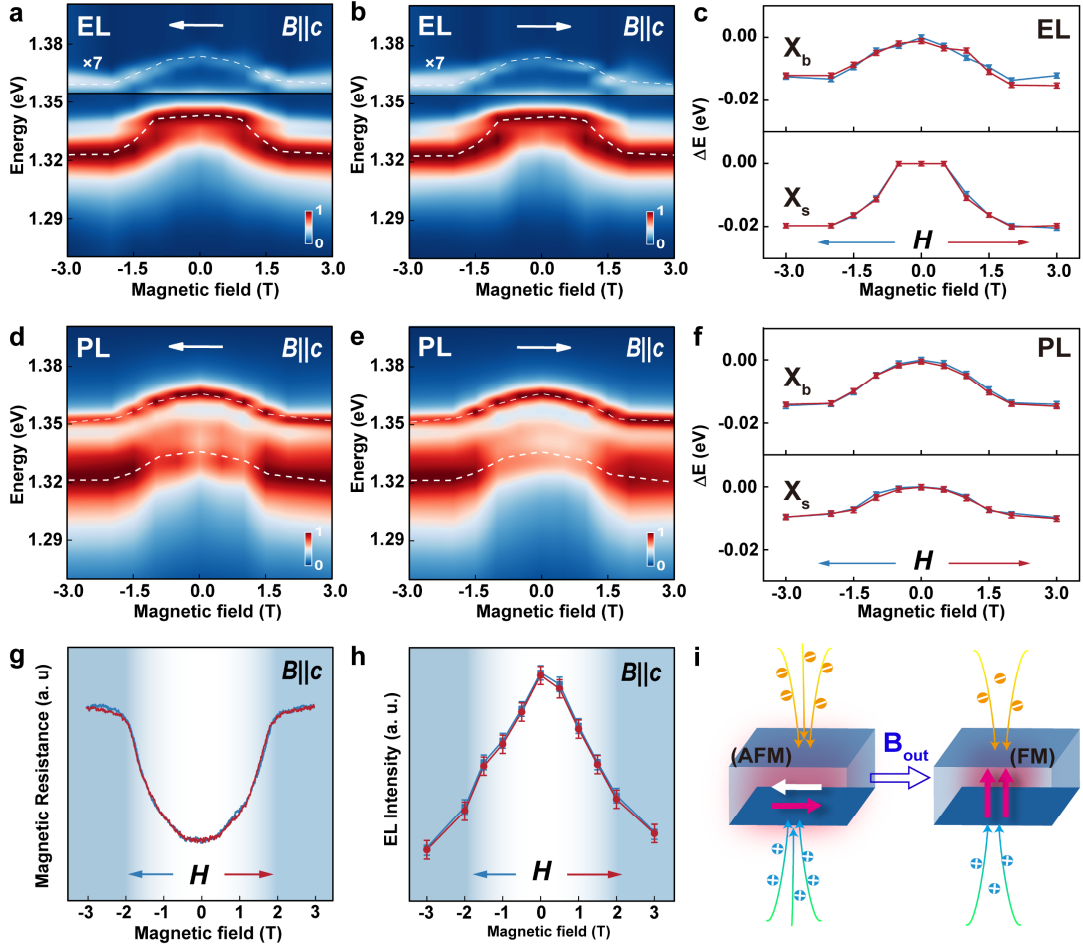

**Fig. 3 Out-of-plane *B* control EL of CrSBr.** Normalized EL spectra of CrSBr as *B* is swept along the hard axis from 3.0 T to -3.0 T (**a**) and from -3.0 T to 3.0 T (**b**). White dashed lines guide the eye to the evolution of $X_s$ and $X_b$ with *B*. (**c**) Calculation of the energy shift difference for the EL spectra during a back-and-forth sweep of *B*, with 0 T referenced to 0 eV. PL spectra of CrSBr as *B* is swept along the easy axis from 3.0 T to -3.0 T (**d**) and from -0.3 T to 0.3 T (**e**). White dashed lines guide the eye to the evolution of $X_s$ and $X_b$ with *B*. (**f**) Calculation of the energy shift difference for the PL spectra during a back-and-forth sweep of *B*, with 0 T referenced to 0 eV. (**g**) MR of CrSBr as *B* is swept along hard axis, with inset showing the out-of-plane-*B* modulated EL emission mechanism. (**h**) EL intensity of CrSBr as *B* is swept along hard axis. (**i**) The mechanism of out-of-plane-*B* modulated EL emission.

Unexpectedly, the EL efficiency during the spin-canting process exhibits an opposite trend with respect to spin-flip transition under the sweeping $B_\text{out}$. Before the saturation field of ~2.0 T, the MR increases during the spin-canting process with increasing magnetic fields (Fig. 3g), while the EL intensities decreases with decreasing MR. This indicates that lower carrier injection efficiency for larger MR, as shown in Fig. 3h. Fig. 3i demonstrates that when $B_{out}$ switches CrSBr from AFM to FM through spin-canting transition, reduced carrier injection efficiency results in weaker EL intensity. Through showing the opposite efficiency trend of EL efficiency in the spin-flip transition, we take similar modulation analysis and quantify the manipulation efficiency of EL by $B_{out}$, which reaches 73.8%, as discussed in detail in Supplementary Information Section IV. Consequently, this device not only provides a novel prototype for electrically driven light emitters in opto-spintronic devices but also represents a significant step towards practical application with the large modulation efficiency.

**Conclusion**

In summary, the prototype spin-LED, as a core scheme of opto-spintronic devices, has been realized and manifests spin-transitions modulated EL down to bilayers of vdW CrSBr. Particularly, the key spectral characteristics of EL exhibit unprecedented hysteresis during the spin-flip transition, but switch to continuous modulations during the spin-canting transition. Besides the spin-order dressed excitonic transitions for the EL spectral resonances, the EL efficiency is directly determined by the spintronic transport, which shows opposite magnetic-modulated trends for spin-flip and spin-canting transitions. This study establishes the prototype device architecture for spin-LED, which essentially merges photonic, electronic, and magnetic functionalities. Moreover, it also promises many opportunities to develop magnetic sensors, magnetic tunneling junctions, integrated 2D opto-spintronic chip for neuromorphic computing.

**Methods**

**Synthesis of CrSBr.**

CrSBr single crystals were synthesized via chemical vapor transport (CVT, Wuhan Shiwei Optoelectronic Technology Co., Ltd.). Chromium (III) bromide ($CrBr_3$, 99.00% purity, Macklin), sulfur powder (S, 99.95% purity, Aladdin), and Chromium powder (Cr, 99.00% purity, thermo scientific) were used as precursors. A stoichiometric mixture of $CrBr_3$ (720 g), S (196 g), and Cr (189 g) was sealed within an evacuated quartz ampoule using an oil-free molecular pump under a high vacuum. The ampoule was subjected to a controlled thermal gradient. The source zone temperature was programmed as follows: ramped to 500 °C over 6 h, then to 550 °C over 12 h, subsequently to 930 °C over 12 h, and finally held at 930 °C for 84 h before cooling naturally to room temperature. Concurrently, the growth zone temperature profile was: ramped to 570 °C over 6 h and held for 12 h, then increased to 970 °C over 12 h and maintained for 24 h, followed by cooling to 850 °C over 12 h and holding for 48 h, before natural cooling to room temperature. The resulting crystals were recovered and stored within an argon-filled glove box upon opening the ampoule.

Thin flakes of CrSBr were prepared using the scotch magic tape-based mechanical exfoliation technique. Prior to exfoliation, $SiO_2$/Si substrates were treated with oxygen plasma for 10 min to remove surface adsorbates and enhance adhesion. CrSBr crystals were then mechanically exfoliated directly onto the treated $SiO_2$/Si substrates. All exfoliation procedures were performed under ambient conditions at room temperature.

**Sample fabrication**

QW devices were fabricated using graphite, hBN, and exfoliated CrSBr flakes. Graphite and hBN crystals were sourced from HQ Graphene. Flake thickness selection for graphite and hBN followed the criteria established in our previous work[50]. Both materials were mechanically exfoliated onto $SiO_2$/Si substrates. Dark-field optical microscopy confirmed the flakes possessed clean surfaces and uniform quality, qualifying them for quantum well assembly. CrSBr flakes of varying layer numbers were initially identified via optical microscopy contrast and subsequently had their thickness verified by atomic force microscopy. Device assembly was performed on a transfer stage. A polydimethylsiloxane (PDMS)/polycarbonate (PC) stamp was used to sequentially pick up the constituent layers (graphite, hBN, CrSBr, hBN, graphite) at 80 °C. The assembled stack was then transferred onto pre-patterned electrodes. These electrodes were fabricated by ultraviolet photolithography followed by thermal evaporation of a 5 nm chromium adhesion layer and a 50 nm gold layer. During the transfer, the PC film was melted onto the electrode structure. Finally, the completed device was immersed in chloroform to dissolve residual PC. This was followed by sequential 5-minute immersions in acetone and isopropanol to ensure the surface remained clean and free of contaminants.

**Experimental setup**

The experimental setup for PL, EL, and electrical characterization incorporated a closed-loop cryofree magneto-optical system (attoDRY 2100) featuring a 9 T superconducting magnet (out-of-plane) and an attocube piezo-driven xyz translation stage. Optical spectra (PL and EL) were collected using a WITec Alpha 300R confocal microscope. PL was excited with a 532 nm laser focused to a spot size of approximately 1 micrometer. Device voltage control was managed by a Keysight B1500A parameter analyzer. AFM imaging in contact mode was conducted on a Bruker DIMENSION ICON instrument employing a DDESP-V2 cantilever (force constant: 80 N/m, resonance frequency: 450 kHz).

**Competing interests**

The authors declare no competing interests.

Correspondence and requests for materials should be addressed to Sheng Liu, Xiaoze Liu or Ting Yu.

**Acknowledgements**

This project was supported by the National Key Research and Development Program of China (No. 2021YFA 1200800), the Start-up Funds of Wuhan University, and the Fundamental Research Funds for the Central Universities (2042024kf0010).

**Author Contributions**

F.L.Q. and H.Y.L. contributed equally to this work. F.L.Q. H.Y.L., S.L., X.Z. L., and T.Y. conceived the project and designed the experiments. Device fabrication and experimental measurements were performed by F.L.Q., H.Y.L., Y.L.L., A.S.Y., X.J.W., Y.S., X.Y.Z., J.Y.Z, and X.L. Data analysis and processing were carried out by F.L.Q., H.Y.L., S.L., X.Z.L., and T.Y. The manuscript was written by F.L.Q. and H.Y.L., with revisions contributed by S.L., X.Z.L., V.L.Z., W.B.G., and T.Y.